\documentclass[aps,pra,twocolumn,showpacs,floatfix,preprintnumbers,amsmath,amssymb]{revtex4}
\usepackage{epsfig}
\usepackage{graphicx}
\usepackage{dcolumn}
\usepackage{bm}

\begin{document}
\title{Classical bifurcations and entanglement in smooth Hamiltonian system}
\author{M. S. Santhanam$^{1}$, V. B. Sheorey$^{1}$ and Arul Lakshminarayan$^{2}$\footnote{Permanent address: Department of Physics, Indian Institute of Technology Madras, Chennai, 600036, India.}}
\affiliation{$^{1}$Physical Research Laboratory, Navrangpura, Ahmedabad 380 009, India.\\
$^{2}$Max Planck Institute for the Physics of Complex Systems,\\
N\"othnitzer Strasse 38., Dresden 01187, Germany}
\date{\today}

\begin{abstract}
We study entanglement in two coupled quartic oscillators.
It is shown that the entanglement, as measured by
the von Neumann entropy, increases with the classical chaos parameter for
generic chaotic eigenstates. We consider certain isolated periodic orbits whose
bifurcation sequence affects a class of quantum eigenstates, called the
channel localized states. For these states, the entanglement is a local minima
 in the vicinity of a pitchfork bifurcation
but is a local maxima near a anti-pitchfork bifurcation. We place these results in 
the context of the close connections that may exist between entanglement measures
and conventional measures of localization that have been much studied in quantum
chaos and elsewhere. We also point to an interesting near-degeneracy that arises in the
spectrum of reduced density matrices of certain states as an interplay of localization
 and symmetry.

\end{abstract}
Preprint Number : IITM/PH/TH/2007/6
\maketitle

\section{Introduction}

The study of entanglement is currently an active area of research in view of
it being a physical resource for
quantum information theory, quantum computing,
quantum cryptography and teleportation \cite{qcomp}. 
At a classical level, entanglement does not have a corresponding counterpart.
However, increasingly it is being realized that the nature of classical
dynamics, whether it is regular or chaotic, affects entanglement in the
quantized version of the system \cite{arul1}.
In general, larger chaos in the system leads to larger entanglement
production. This has been established by considering kicked top
models \cite{ktm}, bakers map \cite{bm}, Dicke model \cite{dicke},
billiard in a magnetic field \cite{bill}, kicked Bose-Einstein condensates \cite{kbec}
and $N$-atom Jaynes-Cummings model \cite{jcm}.
In contrast to these studies, the role of classical bifurcations in
entanglement of chaotic systems has not received much attention.
Even though entanglement is a purely quantum attribute, it is
nevertheless affected by the qualitative nature of the  dynamics in phase
space. The results to this effect are obtained primarily in the context
of quantum phase transitions in the ground state of infinite systems
in which the entanglement is maximal at critical parameter values \cite{entqp}.
For instance, for case of ions driven by laser fields and coupled to a heat bath,
i.e, a form of Dicke model was shown to exhibit maximal entanglement
of its ground state at the parameter value at which classical system bifurcates.
Similar result for the
ground state was reported from the study of coupled tops, a generalization
of the two dimensional transverse field quantum Ising model \cite{hines}
as well from Jahn-Teller models \cite{jtm}.
The ground state entanglement of mono-mode Dicke model
is shown to be related to Hopf bifurcation \cite{nemes}.
Qualitatively similar results for two component Bose-Einstein condensate
are also known \cite{xie}.
In all these cases, the treatment is confined mostly
to the ground state of the system that exhibits criticality
and involves one single classical bifurcation.

Do these results hold good for chaotic, smooth Hamiltonian systems
that do not exhibit criticality in the sense of phase transitions ?
As opposed to a single bifurcation, what happens in bifurcation
sequences where stability loss and stability
gain interleave one another ? Both these questions explore
the connection between chaos and entanglement in a physical setting that is
different from the earlier studies.
In the context of this work,
we examine a Hamiltonian system whose classical dynamics is controlled by
a single tunable parameter. The changes in the parameter leads to
changes in the phase space structure; for instance regularity to chaos
transition and bifurcation sequences of fixed points.
Typically, chaotic systems display a sequence of bifurcations. Consider, for
instance, the coupled oscillator systems, a paradigm of chaos for smooth
Hamiltonian systems and is related to atoms in strong magnetic fields, the
quadratic Zeeman effect problems \cite{wint}.
In these cases,
one particular sequence of bifurcation is a series of pitchfork
and anti-pitchfork bifurcations \cite{delos1}. The pitchfork corresponds to a periodic
orbit losing stability and in the Poincar\`e section this appears as a elliptic
fixed point giving way to a hyperbolic fixed point. The anti-pitchfork
is when the periodic orbit gains stability. In this work, we consider
coupled quartic oscillators and show that the entanglement in the
highly excited states of the system is modulated by classical
bifurcations. We could place this in the context of works that lend support to the notion
that for generic one-particle states there is a strong correlation between entanglement and 
measures  of localization \cite{arulsub,li04,li05}.

\section{Entanglement in a bipartite system}

A pure quantum state $|\Psi\rangle$ composed of many
subsystems $|\phi_i\rangle$ is said to be entangled if it cannot be written down as a
direct product of states corresponding to each of the subsystem.
\begin{equation}
|\Psi\rangle_{\mbox{entangled}} \;\;\; \neq \;\;\; |\phi_1\rangle \otimes |\phi_2\rangle \otimes
|\phi_3\rangle ........ \otimes |\phi_n\rangle
\label{ent_def}
\end{equation}
Thus, entanglement implies stronger than classical correlations.
If $\rho = |\psi\rangle \langle\psi|$ is the density matrix representation for a pure state
$|\psi\rangle$, then the reduced density matrix (RDM) can be obtained by applying
the trace operation to one of the degrees of freedom. Thus,
\begin{equation}
\rho_1 = \mbox{Tr}_2 |\psi\rangle \langle\psi| \;\;\;\;\;\;\;\;\; \rho_2 = \mbox{Tr}_1 |\psi\rangle \langle\psi|
\end{equation}
are two RDMs, whose one of the degrees of freedom is traced out.
The notation $\mbox{Tr}_i$ denotes that the trace operation is applied on the $i$th degree of freedom.
Schmidt decomposition \cite{qcomp} provides a representation for $|\psi\rangle$
in terms of product of basis states,
\begin{equation}
|\psi\rangle = \sum_{i=1}^{N} \sqrt{\lambda_i} \;\; |\phi_i\rangle_{(1)} \;\; |\phi_i\rangle_{(2)}
\end{equation}
where $|\phi_i\rangle_{(1)}$ and $|\phi_i\rangle_{(2)}$ are the eigenvectors
of the RDMs $\rho_1$ and $\rho_2$ respectively, and $\lambda_i$ are the eigenvalues
of either of the RDMs. The von Neumann or the entanglement entropy of pure state is given by,
\begin{equation}
S = - \sum_{i=1}^N \lambda_i \log \lambda_i
\label{vne}
\end{equation}
Thus, when $S=0$, the subsystems are not entangled and when $S>0$, they are
entangled.
The Schmidt decomposition provides a compact and unique representation for the given
eigenstate (unique in the generic case when the non-zero spectrum of the RDM is nondegenerate).

\section{Hamiltonian Model and bifurcation sequence}
\subsection{Quartic oscillator}
We consider the Hamiltonian system given by,
\begin{equation}
H = p_x^2 + p_y^2 + x^4 + y^4 + \alpha x^2 y^2
\label{genham}
\end{equation}
with $\alpha$ being the tunable chaos parameter.
For $\alpha=0,2,6$, the system is classically integrable
and becomes predominantly chaotic as $\alpha \to \infty$.
This has been extensively studied as a model for 
classical and quantum chaos in smooth Hamiltonian
systems \cite{tom} and exhibits qualitatively similar dynamics as the
host of problems involving
atoms in strong external fields. In the limit $\alpha \to \infty$,
it is also of relavance as model of classical Yang-Mills field \cite{ym}.
To study the quantum analogue of this system, we quantize it
in a symmetrized basis set given by,
\begin{equation}
\psi_{n_1,n_2}(x,y) = {\cal N}(n_1,n_2) \left[ \phi_{n_1}(x) \phi_{n_2}(y) + 
\phi_{n_2}(x) \phi_{n_1}(y) \right]
\label{bset}
\end{equation}
where ${\cal N}(n_1,n_2)$ is the normalization constant and $\phi(x) \phi(y)$ is
the eigenstate of unperturbed quartic oscillator with $\alpha=0$.
The choice of this form of basis set is dictated by the fact that the
quartic oscillator has $C_{4v}$ point group symmetry, i.e., all the invariant
transformations of a square. Hence we have chosen the
symmetry adapted basis sets as in Eq. \ref{bset} from $A_1$ representation
of $C_{4v}$ point group.

Thus, the $n$th eigenstate is,
\begin{equation}
\Psi_n(x,y) = \sum_{j(n_1,n_2)=1} a_{n,j(n_1,n_2)} \; \psi_{n_1,n_2}(x,y)
\label{expcoeff}
\end{equation}
where $a_{n,j(n_1,n_2)} = \langle \psi(x,y)|\Psi_n(x,y) \rangle $ are the expansion
coefficients in the unperturbed basis space. Note that $n_1,n_2$ are
even integers and
$a_{n,j(n_1,n_2)}  = a_{n,j(n_2,n_1)}$ in $A_1$ representation
of $C_{4v}$ point group.
The eigenvalue equation is solved numerically by setting up Hamiltonian
matrices of order 12880 using 160 even one-dimensional basis states.

\subsection{Bifurcation sequence in quartic oscillator}
\begin{figure}
\centerline{\includegraphics[height=2.3cm]{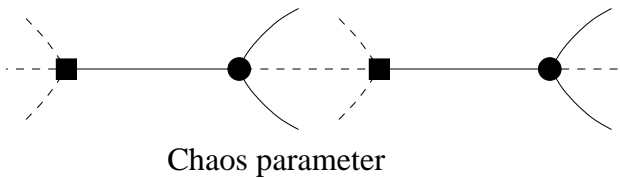}}
\caption{The schematic of a typical bifurcation sequence involving
a series of pitchfork (circles) and anti-pitchfork (square) bifurcations
as a function of chaos parameter.
The solid lines indicate that the orbit is stable and dashed line
indicate instability.}
\label{bifseq}
\end{figure}

In a general chaotic system many bifurcation sequences are possible.
However, a two dimensional Hamiltonian system can exhibit only
five types of bifurcations \cite{delos1}. One such prominent sequence is
a series of pitchfork and anti-pitchfork bifurcation shown schematically
in Fig \ref{bifseq}. To reiterate, a pitchfork bifurcation takes place when a stable
orbit loses stability and gives rise to two stable orbits. Anti-pitchfork
bifurcations happen when a stable orbit is spontaneously born due to
the merger of two unstable orbits.
We will focus on a particular periodic orbit, referred to as the channel
orbit in the literature \cite{mss3}, given by the
initial conditions $\{x,y=0,p_x,p_y=0\}$, which displays
such a bifurcation sequence.
The Poincar\'e section in the vicinity of the channel orbit
has interesting scaling properties and the orbit itself
has profound influence on a series of quantum eigenstates, called
localized states, in
the form of density enhancements or scars \cite{mss2}.
Such density enhancements due to channel orbits have also
been noted in atoms in strong magnetic fields or the
diamagnetic Kepler problem \cite{delos2} as well.

\begin{figure}
\centerline{\includegraphics[height=5cm]{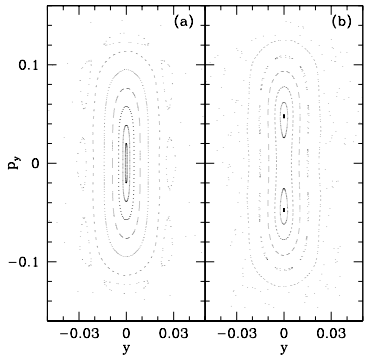}}
\caption{The Poincare section for the quartic oscillator in
Eq. \ref{genham} in shown for (a) $\alpha=90$ and (b) $\alpha=90.5$.
Note that at $\alpha=90$ the periodic orbit
undergoes a pitchfork bifurcation.}
\label{psec90}
\end{figure}

\begin{figure}
\centerline{\includegraphics*[height=4cm]{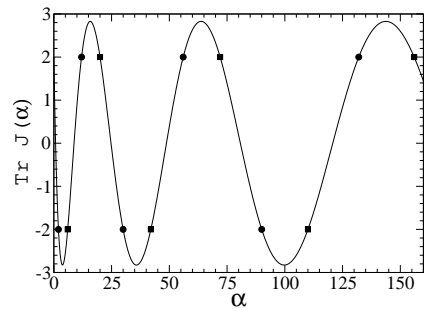}}
\caption{The linear stability of the channel orbit as a function
of $\alpha$. The orbit is stable for $|\mbox{Tr}~ J(\alpha)| < 2$.
The pitchfork bifurcation points are indicated by circles
and anti-pitchfork bifurcations are indicated by squares.}
\label{orb-sta}
\end{figure}

The stability of the channel orbit in the quartic oscillator
in Eq (\ref{genham}) is indicated by the trace of monodromy
matrix $J(\alpha)$ obtained from linear stability analysis.
It can be analytically obtained for the channel orbits \cite{yosh}  as,
\begin{equation}
\mbox{Tr}~ J(\alpha) = 2 \sqrt{2} \cos\left( \frac{\pi}{4} \sqrt{1+4\alpha}\right).
\label{trj}
\end{equation}
The channel orbit is stable as long as $|\mbox{Tr} J(\alpha)| < 2$ and it
undergoes bifurcations whenever $\mbox{Tr} J(\alpha) = \pm 2$. From this condition,
it is clear that the bifurcations take place at $\alpha_n = n(n+1)$, $(n=1,2,3....)$.
Thus the channel orbit undergoes an infinite sequence of pitchfork and anti-pitchfork
bifurcations at $\alpha=\alpha_n$. Note that for $n=9$, we have $\alpha = 90$ as one
of the pitchfork bifurcation points. The Poincar\'e sections displayed in
Fig \ref{psec90} shows
that the stable channel orbit at $\alpha=90$ (Fig \ref{psec90}(a)) bifurcates
and gives birth to two new stable orbits (Fig \ref{psec90}(b)) while the channel
orbit itself becomes unstable. Thus, pitchfork bifurcations take place at
$\alpha_n = 2, 12, 30, 56, 90, .....$ and anti-pitchfork at
$\alpha_n = 6, 20, 42, 72, ....$. This can be observed in the plot of
$\mbox{Tr} J(\alpha)$ as a function of $\alpha$ shown in Fig \ref{orb-sta}.

\section{Quartic oscillator states and reduced density matrix}
\subsection{Quartic oscillator spectra}
\begin{figure}
\centerline{\includegraphics*[height=5cm]{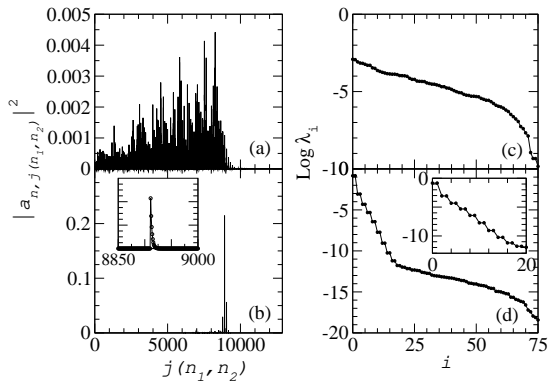}}
\caption{Quartic oscillator eigenstates for $\alpha=90$ in the unperturbed
basis. (a) 1973rd state (delocalized), (b) 1972nd state (localized). The inset
is the magnification of the dominant peak. The
eigenvalues of the RDMs for (c) 1973rd state and (d) 1972nd state. The
inset in (d) is the magnification of the dominant eigenvalues that display
degeracy.}
\label{eigs}
\end{figure}

The quantum spectrum of the quartic oscillator is extensively
studied and reported \cite{tom,mss2,mss1}. For the purposes of this study, we note that
two classes of eigenstates can be identified. The first one is
what we call a generic state whose probability density $|\Psi_n(x,y)|^2$
covers the entire accessible configuration space. Most of the
eigenstates fall in this class and they are instances of
Berry's hypothesis that the Wigner function of a typical chaotic
state condenses on the energy shell \cite{berry1}. In Fig \ref{eigs}(a), we show
the expansion coefficients for the 1973rd eigenstate of the quartic
oscillator
counted sequentially from the ground state for $\alpha=90$. Notice that
the state is delocalized over a large set of basis states. These class
of states are well described by random matrix theory.
The second class of states
is the localized states, which has enhanced probability density
in the vicinity of the underlying classical periodic orbits.
Theoretical support for this class of states based on semiclassical
arguments is obtained from the works of Heller \cite{heller}, Bogomolny and Berry \cite{bob}. As a typical
case, Fig \ref{eigs}(b) shows the expansion coefficients for the
1972nd state which is localized over very few basis states in contrast
to the one in Fig \ref{eigs}(a).
In this work, we concentrate on a subset of such eigenstates
whose probability density is concentrated in the vicinity of the
channel periodic orbit. This set of states are nearly separable and
can be approximately labelled by a doublet of quantum numbers $(N,0)$
using the framework of adiabatic theory \cite{mss2,mss1}. Note that such a labeling
is not possible for the generic states since they are spread over a large
number of basis states.

\subsection{Reduced density matrix}
In this section, we compute the eigenvalues of the RDM and
the entanglement entropy of the
quartic oscillator eigenstates as a function of the chaos parameter
$\alpha$. In terms of the expansion coefficients in Eq. (\ref{expcoeff}),
the elements of RDM, ${\mathbf R_x}$, can be written down as,
\begin{equation}
\langle n_2|\rho^{(x)}|n_2' \rangle = \sum_{n_1=1}^M K_{n_1,n_2}a_{n_1,n_2} a_{n_1,n_2'},
\end{equation}
where the normalization constant $K_{n_1,n_2}=1$ if $n_1=n_2$ and $=1/2$ if
$n_1 \ne n_2$. In this case, the $y$-subsystem has been traced out.
Similarly another RDM, ${\mathbf R_y}$, with elements
$\langle n_2|\rho^{(y)}|n_2' \rangle$ can be obtained by tracing over
$x$ variables.
Let $\mathbf A$ represent the eigenvector matrix of order $(M+2)/2$ with
elements $a_{n_1,n_2}$, where $n_1, n_2 = 0,2,4 ....M$
labels the rows and columns respectively. Then, in matrix language,
the RDM ${\mathbf R_x} = \mathbf{A^T A}$ is matrix of order $(M+2)/2$.

In our case, $M=318$ and we numerically diagonalize the RDM of order 160.
The eigenvalues of RDM for a typical delocalized state and a localized state
is plotted in Fig \ref{eigs}(c,d). In general, the dominant eigenvalues
fall exponentially, though with different rates, for both the generic and 
typical localized state
indicating that the Schmidt decomposition provides a compact
representation for the given eigenstate. Earlier such a behavior was noted
for coupled standard maps \cite{arul1}.
The first few dominant eigenvalues of RDM for localized states
display (near-)degeneracy (see Fig \ref{eigs}(d)). This arises
as a consequence of ($i$) $C_{4v}$ symmetry of the potential due to which
the eigenvector matrix is symmetric, i.e,  $a_{n_1,n_2} = a_{n_2,n_1}$ and
($ii$) the localization is exponential in the direction
perpendicular to that in which the quanta of excitation is larger \cite{mss1},
i.e, $a_{N,n_2} \propto \exp(-\omega n_2)$, where $\omega > 0$ is a constant
independent of $N$.

The origin of near-degeneracy can be understood by
by considering a simple model of $4 \times 4$
symmetric eigenvector matrix (the state number index $n$ is suppressed
such that $a_{n,j(n_1,n_2)} =a_{n_1,n_2}$),
\begin{equation}
\mathbf{P} = \left( \begin{array}{cccc}
a_{0,0} & a_{2,0} & a_{N,0} & a_{N+2,0} \\
a_{2,0} & a_{2,2} & a_{N,2} & a_{N+2,2} \\
a_{N,0} & a_{N,2} & a_{N,N} & a_{N+2,N} \\
a_{N+2,0} & a_{N+2,2} & a_{N+2,N} & a_{N+2,N+2} \\
\end{array}  \right).
\label{appev1}
\end{equation}
Here we have only used the one-dimensional quartic oscillator
quantum numbers $(0,2,N,N+2)$ because
the localized states can be approximately well represented by
all possible  doublets arising from these quantum numbers. For instance,
an adiabatic separation with the $(N,0)$ manifold gives a good
estimate for the energy of its localized states \cite{mss1}.
The representation gets better as we add more 1D quantum numbers to
the list above.
The exponential localization
implies that $a_{n_1,n_2} \approx 0$ for $n_1 \sim n_2$. Further, $a_{n_1,n_2} \approx 0$
if $n_1,n_2 << N$. Thus, elements $a_{N,N} \sim a_{N+2,N} \sim a_{N+2,N+2} \sim a_{0,0} \approx 0$.
Then, we can identify a block matrix $B$ with non-zero elements as,
\begin{equation}
\mathbf{B} = \left( \begin{array}{cc}
a_{N,0} & a_{N+2,0} \\
a_{N,2} & a_{N+2,2} \\
\end{array}  \right)
\label{appev2}
\end{equation}
Then, the eigenvector matrix $\mathbf P$ can be approximated as,
\begin{equation}
\mathbf{P} \approx \left( \begin{array}{cc}
\mathbf 0 & \mathbf{B} \\
\mathbf{B^T} & \mathbf 0 \\
\end{array}  \right).
\label{appev4}
\end{equation}
Under the conditions assumed above, the RDM separates into two
blocks which are transpose of one another.
Thus, the reduced density matrix will have the form,
\begin{equation}
\mathbf R = \mathbf{P^T P} = \left( \begin{array}{cc}
\mathbf{B B^T} & \mathbf 0 \\
\mathbf 0 & \mathbf{B^T B}\\
\end{array}  \right)
\label{appev3}
\end{equation}
Since the eigenvalues remain invariant under transposition of a matrix, {\it i.e},
the eigenvalues of $\mathbf{B B^T}$ and $\mathbf{B^T B}$ are identical and hence
we obtain the degeneracy.
Though we use a $4 \times 4$ matrix to illustrate the idea, this
near degeneracy would arise for any eigenvector matrix of even order,
if the symmetry and exponential decay conditions are satisfied.

For the localized state shown in Fig \ref{eigs}(b), $N=264$ and
the dominant eigenvalue of RDM
using the approximate scheme in Eqns (\ref{appev1}-\ref{appev3}),
is $\lambda_1=0.4434$. This
is doubly degenerate and compares favorably with the exact numerical
result of 0.4329.
As observed in Fig \ref{eigs}(d), the degeneracy breaks down as we travel down
the index. As pointed out, the dominant eigenvalues of RDM correspond
to definite 1D quantum oscillator modes that exhibit exponential decay
in the perpendicular mode. This is not true of all the oscillator
modes and hence the degeneracy is broken.

\subsection{Entanglement entropy}
\begin{figure}
\centerline{\includegraphics*[height=5cm]{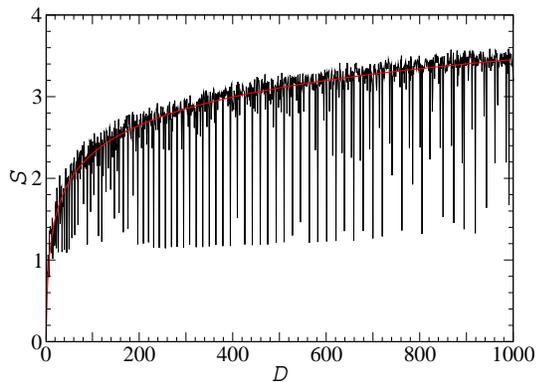}}
\caption{(Color Online) Entanglement entropy for the quartic oscillator at $\alpha=30$
from ground state to 1000th state. The localized states have lower value
of entanglement entropy as seen from the dips in the curve. The solid red curve
is $S_{RMT}$, the RMT average of entanglement entropy.}
\label{vne30}
\end{figure}

Entanglement entropy for each eigenstate is computed from the eigenvalues
of the RDM using Eq (\ref{vne}). In Fig \ref{vne30}, we show the
entanglement entropy of the quartic oscillator at $\alpha=30$ for one 
thousand eigenstates starting from the ground state. The localized
states have values of entanglement entropy much lower than the local average
as seen from the dips in the figure. Most of them are much
closer to zero and substantiate the fact that they are nearly
separable states. In the next section we will show that the entanglement
entropy of localized state is modulated by the bifurcation in the underlying
channel periodic orbit.

The generic delocalized states, on the other hand, form the
background envelope seen in Fig \ref{vne30}. These  chaotic states are not affected
by the bifurcations in the isolated orbits. It is known that such
delocalized states can be modeled using random matrix theory and hence
the distribution of their eigenvectors follows Porter-Thomas distribution \cite{ptd}.
The entanglement entropy can also be calculated based on RMT assumptions
and it is known to be \cite{jay}, $S_{RMT} = \ln(\gamma M)$
where $\gamma \approx 1/\sqrt{e}$ and $M$ is the dimensionality of the
reduced density matrix. In the case of quartic oscillator, the Hilbert space
is infinite in dimension and we take $M$ to be the effective dimension $M_{eff}$
of the RDM. One indicator of the effective dimension of the state is the inverse
participation ratio of the eigenstates. Based on this measure and due
to $C_{4v}$ symmetry of the quartic oscillator, we have $M_{eff}^2=D$ where
$D$ is the state number. Thus, the effective dimension of RDM is, $M_{eff}=\sqrt{D}$.
Finally, we get for the entanglement entropy,
\begin{equation}
S_{RMT} = \ln (\gamma M_{eff} ) \sim \ln (\gamma \sqrt{D}).
\end{equation}
In Fig \ref{vne30}, $S_{RMT}$ is shown as solid red curve and it correctly
reproduces the envelope formed by the delocalized states while the
localized states stand out as deviations from RMT based result, namely,
$S_{RMT}$.

\section{Entanglement entropy and bifurcations}
\begin{figure}
\centerline{\includegraphics*[height=5cm]{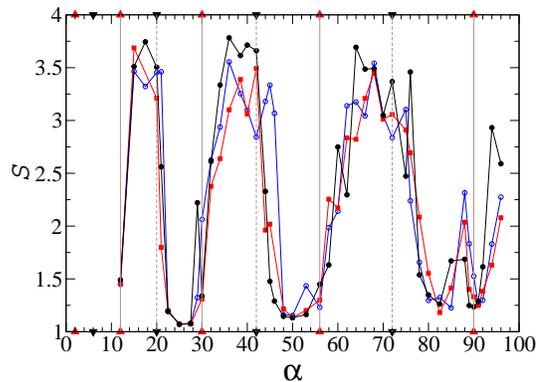}}
\caption{(Color Online) Entanglement entropy as a function of  $\alpha$.
The three curves correspond to different $(N,0)$ type localized states;
solid circles (240,0), open circles (200,0) and squares (210,0). The positions
of pitchfork bifurcations (triangle up) and anti-pitchfork bifurcations
(triangle down) are marked on both the $x$-axes.}
\label{vne_alpha}
\end{figure}
 In this section, we show the central result of the paper that
the entanglement entropy is a minimum at the points at which
the underlying periodic orbit undergoes a pitchfork bifurcation.
As pointed out before, the localized states of the quartic oscillator
are characterized by the doublet $(N,0)$ and are influenced by the
channel periodic orbit. We choose a given localized state,
say, with $N=200$ and compute the entanglement of the same state, i.e,
$(200,0)$ state as a function of $\alpha$. The state that can be characterized
by the doublet $(200,0)$ will be a localized state at every value of
$\alpha$.
The result is shown in Fig \ref{vne_alpha} as the curve plotted
with open circles. The values of $\alpha$ at which the pitchfork and
anti-pitchfork bifurcation takes is marked in both the horizontal axes of
the figure as triangle-up and triangle-down respectively. For the purpose
of easier visualization, they are connected by vertical lines.
Notice that the entanglement entropy attains a local minima
in the vicinity of every classical pitchfork bifurcation and it
attains a local maxima near every anti-pitchfork bifurcation. As
Fig \ref{vne_alpha} shows, similar
result is obtained for two different localized states with
$(N=210,0)$ and $(N=240,0)$. All these localized states are in the
energy regime of highly excited states where the classical system
is predominantly chaotic. The striking similarity between the classical 
stability curve for a particular periodic orbit, namely the channel orbit,
in Fig.~(\ref{orb-sta}) and 
the variation of the entanglement of the localized state is to be noted.
We have also numerically verified (not shown here) that a similar result is obtained
 in the case of another potential where
pitchfork and anti-pitchfork bifurcations of the channel periodic
orbit play an important role,
namely, in $V(x,y)=x^2 + y^2 + \beta x^2 y^2$, where $\beta$ is the
chaos parameter.

At a pitchfork bifurcation, as shown in Fig. \ref{psec90}, the fixed point
corresponding to the channel periodic orbit loses stability
and becomes a hyperbolic point. The central elliptic island seen in
Fig \ref{psec90}(a), breaks up into two islands. The localized state that
mainly derives its support from the classical structures
surrounding the stable fixed point suffers some amount of
delocalization, but is largely supported by the stable regions.
At an anti-pitchfork
bifurcation, the hyperbolic point becomes an elliptic fixed point
and the orbit has gained stability and
a small elliptic island just comes into
existence. Hence, the eigenstate is still largely delocalized since
the small elliptic island is insufficient to support it. This heuristic
picture which is quite sufficient to explain oscillations in localization 
measures is seen to be surprisingly valid even for the somewhat less
intuitive measure of entanglement.
It has been noted earlier 
that when the corresponding classical system undergoes a
pitchfork bifurcation, the entanglement entropy defined
by Eq (\ref{vne}) attains a {\it maximum} \cite{hines}, and it has been
conjectured to be a generic property. We note that this, apparently contradictory result, is 
however in the context of an equilibrium point undergoing a
bifurcation and the relevant state is the ground state, whereas 
in the case we are studying here the orbit that is bifurcating is a
periodic orbit and the states are all highly excited. In this situation there
is a much tighter correlation between more conventional measures
of  localization (Shannon entropy, participation ratio etc.) and
entanglement. 

As the parameter $\alpha$ is increased, the quartic oscillator gets
to be predominantly chaotic and this should imply increase in entanglement.
However, this is true only for the generic delocalized states as
seen in Fig \ref{vne30}. The localized states are influenced not so much
by the increasing volume of chaotic sea but by the specific periodic
orbits that underlie them. Hence, for these states, it is only to be
expected that the qualitative changes in the phase space in the vicinity
of the corresponding periodic orbits affect quantum eigenstate and
hence its entanglement as well. This can be expected to be a generic
feature of entanglement in quantum eigenstates of mixed systems.

\section{Conclusions}
In summary, we considered a smooth Hamiltonian, namely the
two-dimensional, coupled quartic oscillator as a bipartite system. We study
the effect of
classical bifurcations on the entanglement of its quantum eigenstates.
The quartic oscillator is a classically chaotic system. One particular
class of eigenstates of the quartic oscillator, the localized states are scarred by
the channel periodic orbits. We have shown that the entanglement entropy
of these localized states is modulated by the bifurcations in the
underlying channel periodic orbit. When this orbit undergoes a pitchfork bifurcation,
the entanglement attains a local minimum and iwhen it undergoes an anti-pitchfork
bifurcation the entanglement is a local maximum. Physically, this is related
to the presence or the absence of elliptic islands in the phase space
in the vicinity of the channel orbit. We expect this to be a general
feature of bipartite quantum systems whose classical analogue display
bifurcation features.

\acknowledgements We thank J. N. Bandyopadhyay for discussions and comments.

\end{document}